\documentclass[%
 reprint,
superscriptaddress,
 amsmath,amssymb,
 aps
pra,
longbibliography
]{revtex4-1}
\usepackage[dvipsnames]{xcolor}
\usepackage[english]{babel}

\makeatletter
\def\bbl@set@language#1{%
  \edef\languagename{%
    \ifnum\escapechar=\expandafter`\string#1\@empty
    \else\string#1\@empty\fi}%
  \@ifundefined{babel@language@alias@\languagename}{}{%
    \edef\languagename{\@nameuse{babel@language@alias@\languagename}}%
  }%
  \select@language{\languagename}%
  \expandafter\ifx\csname date\languagename\endcsname\relax\else
    \if@filesw
      \protected@write\@auxout{}{\string\select@language{\languagename}}%
      \bbl@for\bbl@tempa\BabelContentsFiles{%
        \addtocontents{\bbl@tempa}{\xstring\select@language{\languagename}}}%
      \bbl@usehooks{write}{}%
    \fi
  \fi}
\newcommand{\DeclareLanguageAlias}[2]{%
  \global\@namedef{babel@language@alias@#1}{#2}%
}
\makeatother

\DeclareLanguageAlias{en}{english}

\usepackage{graphicx}% Include figure files
\usepackage{amsmath}
\usepackage{ulem}
\usepackage{cancel}
\usepackage{array}
\usepackage{amssymb}
\usepackage{dcolumn}% Align table columns on decimal point
\usepackage{bm}% bold math
\let\savecorresponds\corresponds
\let\corresponds\relax
\usepackage{mathabx}
\usepackage{enumitem} 
\usepackage{tabularx}
\usepackage{soul}
\let\corresponds\savecorresponds
\renewcommand{\vec}[1]{\mathbf{#1}}

\def\vec#1{\boldsymbol{#1}}

\def\pd2v#1#2#3{\frac{\partial^2 #1}{\partial #2 \partial #3}}

\def \vec#1{\mathbf{#1}}
\def \2x2mat#1#2#3#4{
\left( \begin{array}{cc}
#1 &  #2 \\  #3 &  #4
\end{array} \right)
}

\begin{document}

\preprint{APS/123-QED}

\title{Quantum image distillation}% Force line breaks with \\

\author{Hugo Defienne}
\affiliation{School of Physics and Astronomy, University of Glasgow, Glasgow G12 8QQ, UK\
}%
\author{Matthew Reichert}%
\affiliation{
Department of Electrical Engineering, Princeton University, Princeton, NJ 08544, USA\
}%
\author{Jason W. Fleischer}%
\affiliation{
Department of Electrical Engineering, Princeton University, Princeton, NJ 08544, USA\
}%
\author{Daniele Faccio}%
\affiliation{School of Physics and Astronomy, University of Glasgow, Glasgow G12 8QQ, UK\
}%

\date{\today}

\begin{abstract}
Imaging with quantum states of light promises advantages over classical approaches in
terms of resolution, signal-to-noise ratio and sensitivity. However, quantum detectors are particularly sensitive sources of classical noise that can reduce or cancel any quantum advantage in the final result. Without operating in the single-photon counting regime, we experimentally demonstrate distillation of a quantum image from measured data composed of a superposition of both quantum and classical light. We measure the image of an object formed under quantum illumination (correlated photons) that is mixed with another image produced by classical light (uncorrelated photons) with the same spectrum and polarisation and we demonstrate near-perfect separation of the two superimposed images by intensity correlation measurements. This work provides a novel approach to mix and distinguish information carried by quantum and classical light, which may be useful for quantum imaging, communications, and security.
\end{abstract}

\maketitle

Quantum imaging exploits photon correlations to overcome fundamental limits of classical imaging. Spatial correlations between pairs of photons are particularly attractive due to their natural high-dimensional structure~\cite{howell_realization_2004,salakhutdinov_full-field_2012,krenn_generation_2014} and the simplicity of photon pair generation from spontaneous parametric down conversion (SPDC)~\cite{malygin_spatiotemporal_1985}. Demonstrations using spatially entangled photon pairs range from ghost imaging~\cite{pittman_optical_1995} to sub-shot-noise imaging~\cite{jedrkiewicz_detection_2004, brida_experimental_2010} and enhanced-resolution imaging~\cite{xu_experimental_2015}. In recent years, important progress has been made in quantum light detection to develop applications from these proof-of-principle experiments. In that regard, multi-pixel single-photon sensitive cameras, such as thresholded electron multiplied charge coupled device (EMCCD)~\cite{lantz_multi-imaging_2008} and single-photon avalanche photodiode (SPAD) cameras~\cite{charbon_spad-based_2013}, have demonstrated great potential to perform high-dimensional coincidence measurements for entanglement characterization~\cite{moreau_realization_2012,tasca_imaging_2012,unternahrer_coincidence_2016}, sub-Rayleigh imaging~\cite{guerrieri_sub-rayleigh_2010} and super-resolution microscopy~\cite{schwartz_superresolution_2013, antolovic_SPAD_2017}.
However, all these quantum detectors operate in the single-photon counting regime (i.e. photons detected one by one), making them extremely vulnerable to sources of classical noise (e.g. background illumination, spurious reflection, etc). For example, an excess of spurious photons detected in a SPAD-based quantum imaging system~\cite{tenne_super-resolution_2018} is likely to saturate the sensor and severely hinder its use. To date, there is still no obvious means of distinguishing a quantum image from classical noise or from a superimposed classical image. Moreover, this problem extends beyond imaging and is tightly related to quantum-classical information discrimination in communications and cryptography~\cite{lydersen_hacking_2010}.\\ 
\begin{figure}
\includegraphics[width=1 \columnwidth]{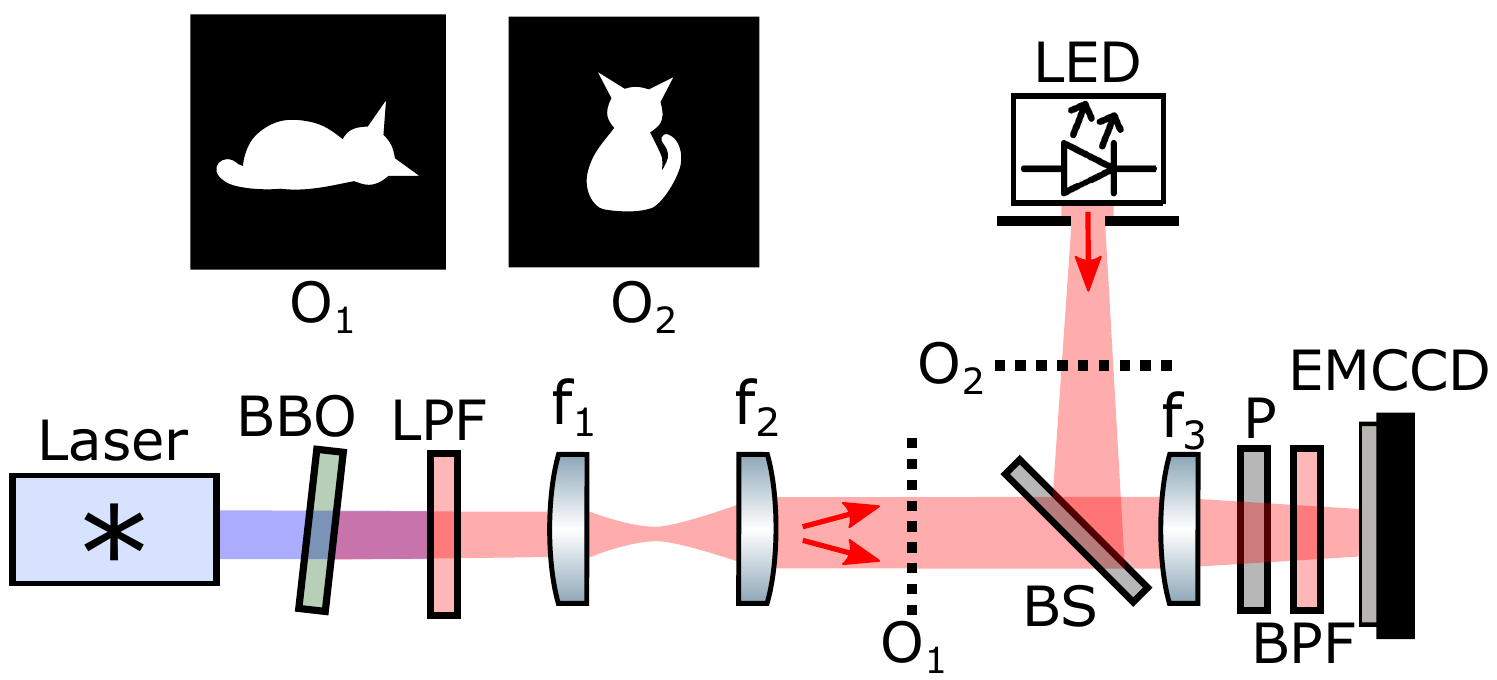}% Here is how to import EPS art
\caption{\label{Figure1} \textbf{Experimental apparatus.}  Light emitted by a diode laser ($\lambda_p=405nm$) illuminates $\beta$-Barium Borate (BBO) crystal of 0.5 mm thickness to produce spatially entangled pairs of photons by type-I SPDC. Long-pass filters (LPF) positioned after the crystal remove pump photons. Lenses $f_1 =35$mm and $f_2 = 75$mm image the crystal surface onto an object $O_1$ (`dead cat'). Simultaneously, an object $O_2$ (`alive cat') is illuminated by a spatially filtered light-emitting diode (LED). Images of both objects are superimposed onto an electron-multiplied charge-coupled device (EMCCD) camera using a single-lens imaging configuration ($f_3 = 50$mm) and an unbalanced beam splitter (92\% transmission) (BS). Band-pass filters (BPF) at $810 \pm 5$nm and a polariser (P) in front of the camera select near-degenerate photons. The single and double red arrows indicate respectively classical and photon pairs illuminations.}
\end{figure}
\begin{figure*}
\includegraphics[width=0.89 \textwidth]{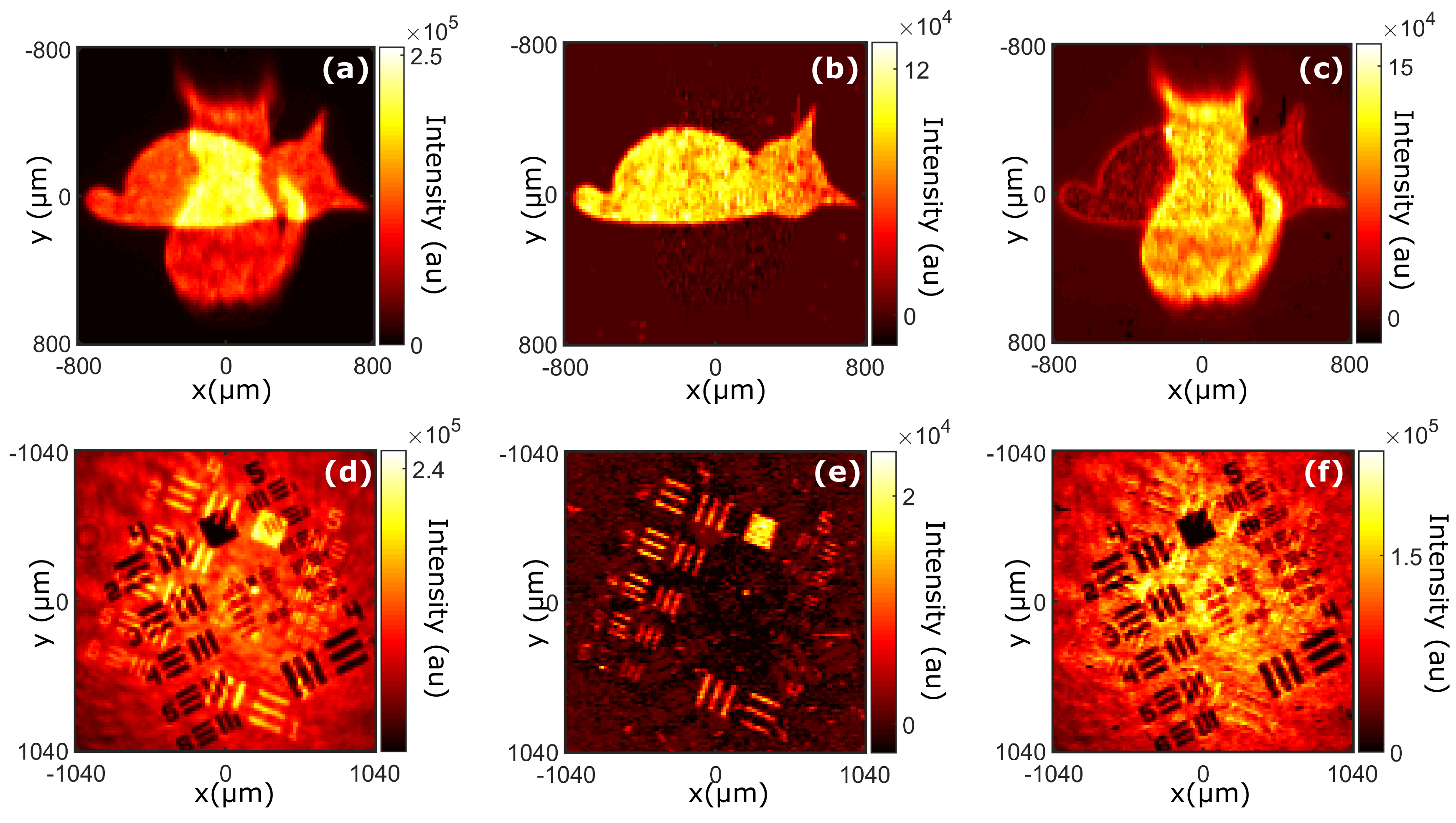}% Here is how to import EPS art
\caption{\label{Figure2} \textbf{Separation of mixed quantum-classical images.} The direct-intensity image \textbf{(a)} acquired by accumulating photons on the camera sensor shows a superposition of both objects $O_1$ (quantum) and $O_2$ (classical), representing respectively a `dead' and an `alive' cat. Intensity correlation function $\Gamma(\vec{r},\vec{r})$ \textbf{(b)} measured with the camera shows the image of $O_1$ . An image of $O_2$ \textbf{(c)} is obtained by subtracting the reconstructed image of $O_1$ from the mixed image. The residual image of $O_1$ observed in the background is due to single photons created by absorption of one photon of a pair propagating through the `dead cat' mask. A similar experiment is performed using positive ($O_1$) and negative ($O_2$) resolution charts, as shown by its corresponding \textbf{(d)} direct-intensity image, \textbf{(e)} $\Gamma(\vec{r},\vec{r})$ and \textbf{(f)} reconstructed classical image. Both experiments are performed by acquiring $N \sim 10^7$ frames using an exposure time of $\tau = 6$ms.}
\end{figure*}
In this letter, we report an experimental technique that allows the distillation of a quantum image from a camera measurement that contains both a quantum and a classical image. No prior information of the images themselves is required other than the statistics of the illuminating sources (i.e. the quantum image is encoded in correlated photon-pair events). An object illuminated by spatially entangled photon pairs forms an image that is mixed with that of another object illuminated by classical coherent light. Both images are indistinguishable in terms of spectrum and polarization, so that conventional intensity measurements cannot discern between them. However, intensity correlation measurements are sensitive to photon statistics. While photons emitted by the classical coherent source are uncorrelated~\cite{glauber_quantum_1963}, pairs of photons in the SPDC illumination are correlated in position~\cite{walborn_spatial_2010,schneeloch_introduction_2016}. We exploit these spatial intensity correlations to extract an image of the object illuminated by photon-pairs from a mixed quantum-classical image and thus reconstruct the classical image by subtraction. We finally investigate the impact of classical light on the signal-to-noise ratio and show that quantum information can be retrieved even when the classical illumination is ten times higher than the quantum illumination.\\
\begin{figure*}
\includegraphics[width=0.8 \textwidth]{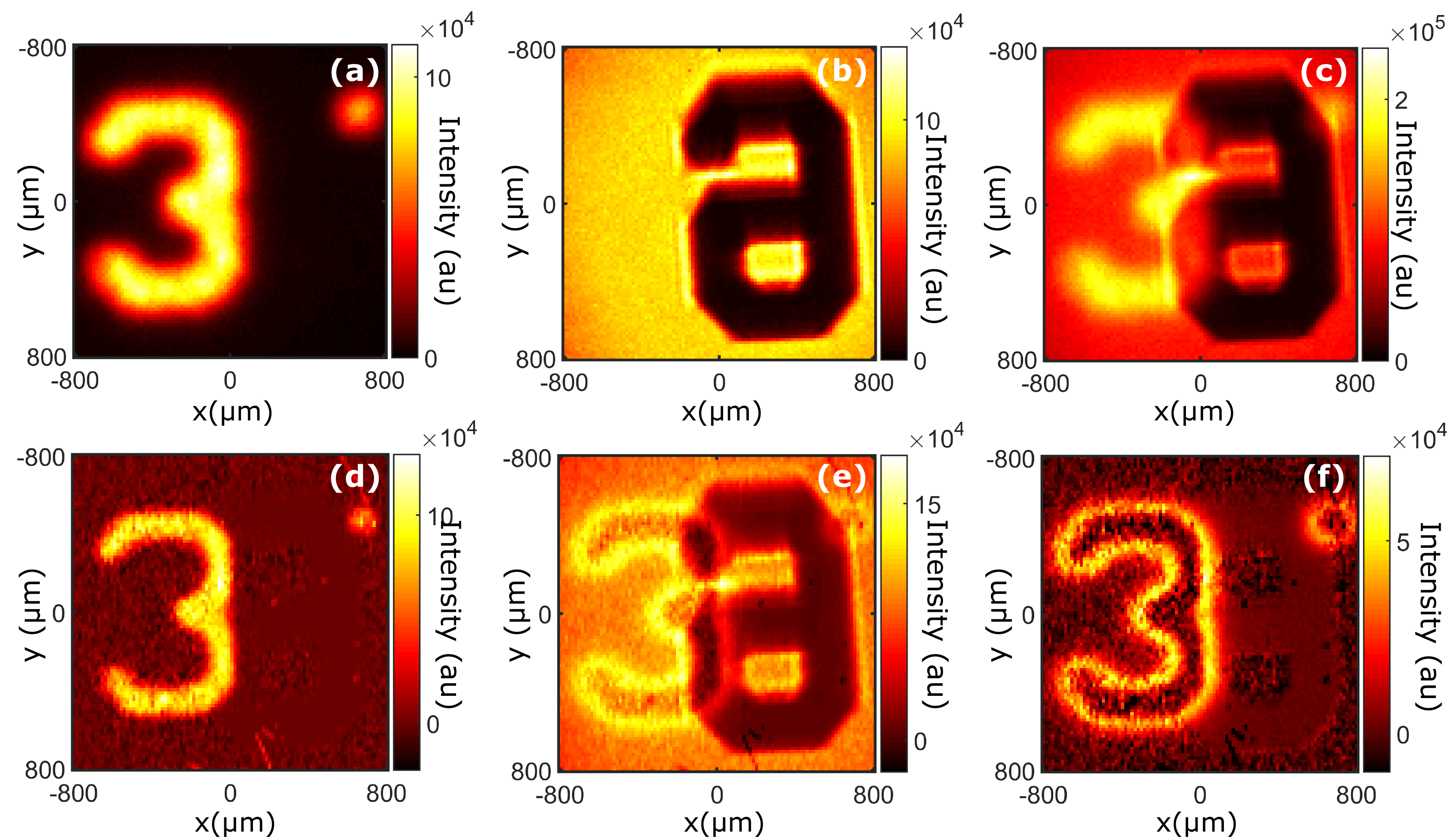}% Here is how to import EPS art
\caption{\label{Figure3} \textbf{Characterization of residual single-photon intensity.} Direct-intensity image \textbf{(a)} acquired with the LED turned off shows object $O_3$ (the number `3'). The image is deliberately slightly defocused by positioning it out of the focal plane of the imaging system. Direct-intensity image \textbf{(b)} acquired with the SPDC turned off shows the ground-truth image of $O_4$ (the number `6'). Direct-intensity image \textbf{(c)} acquired with both sources on shows a superimposition of both objects. The intensity correlation function $\Gamma(\vec{r},\vec{r})$ \textbf{(d)} reveals the number `3'; image subtraction between this  and the mixed image reveals the classical image \textbf{(e)}, number `6'. In this case, the residual intensity created by absorption of one photon of a pair is concentrated near the edge of the number `3'. The residual single-photon intensity \textbf{(f)} is isolated by subtracting the reconstructed classical \textbf{(e)} from its ground truth \textbf{(b)}. Experiments are performed by acquiring $N=6.10^6$ frames using an exposure time of $\tau = 6$ms.  }
\end{figure*}
Figure~\ref{Figure1} shows the experimental setup. A collimated laser beam ($ 405$nm) interacts with a tilted non-linear crystal of $\beta$-barium borate (BBO) to produce pairs of infrared photons by type-I SPDC. The down-converted field at the output of the crystal is imaged onto an object $O_1$ (`dead cat') using a two-lens imaging system $f_1-f_2$. Simultaneously, a spatially filtered light emitting diode (LED) illuminates a second object $O_2$ (`alive cat'). A single-lens imaging system ($f_3$) and an unbalanced beam splitter (92\% transmission) image both objects onto an EMCCD camera. Narrowband-pass filters (BPF) and polarisers (P) ensure that all photons falling on the camera sensor have the same wavelength ($810 \pm 5$ nm) and polarization. 

Figure~\ref{Figure2}(a) shows an intensity image acquired by photon accumulation on the camera under simultaneous illumination from both sources. Objects $O_1$ and $O_2$ (i.e. both the `dead' and `alive' cats) are superimposed. Figure~\ref{Figure2}(b) shows an image of $\Gamma(\vec{r},\vec{r})$, where $\Gamma$ is the intensity correlation function and $\vec{r}$ is a camera pixel position. As detailed in Methods, $\Gamma$ is retrieved using the full dynamic range of the camera (i.e. no photon-counting) which prevents the sensor from saturating due to multiple photon detections. Remarkably, only the object $O_1$ that is illuminated by down-converted light (i.e. the `dead cat') is apparent. Since photons emitted by the classical source are uncorrelated, the only non-null contribution to $\Gamma$ is due to entangled photon pairs produced by SPDC. When pairs of photons correlated in position illuminate homogeneously an object $O_1$~\cite{abouraddy_entangled-photon_2002}, $\Gamma(\vec{r},\vec{r})$ is proportional to its shape.
\begin{equation}
\Gamma(\vec{r},\vec{r}) \sim |O_1(\vec{r})|^4
\end{equation} 
Not only does this approach allow near-perfect reconstruction of the quantum image, but it also enables to retrieve the classical image (i.e. `alive cat') by subtracting the quantum image [Fig.~\ref{Figure2}(b)] from the mixed image [Fig.~\ref{Figure2}(a)], as shown in Fig.~\ref{Figure2}(c). The same experiment performed with more complex objects (i.e. resolution charts in Fig.~\ref{Figure2}d) continues to show a very good extraction of the quantum image [Fig.~\ref{Figure2}e]. However, we observe the presence of residual intensities in the retrieved classical images [Fig.~\ref{Figure2}c and f] that are located near the edges and in the head of the `dead cat' mask. This effect is due to single photons created by absorption of one photon of a pair when propagating through the objects~\cite{reichert_biphoton_2017}.\\
\begin{figure}
\includegraphics[width=1 \columnwidth]{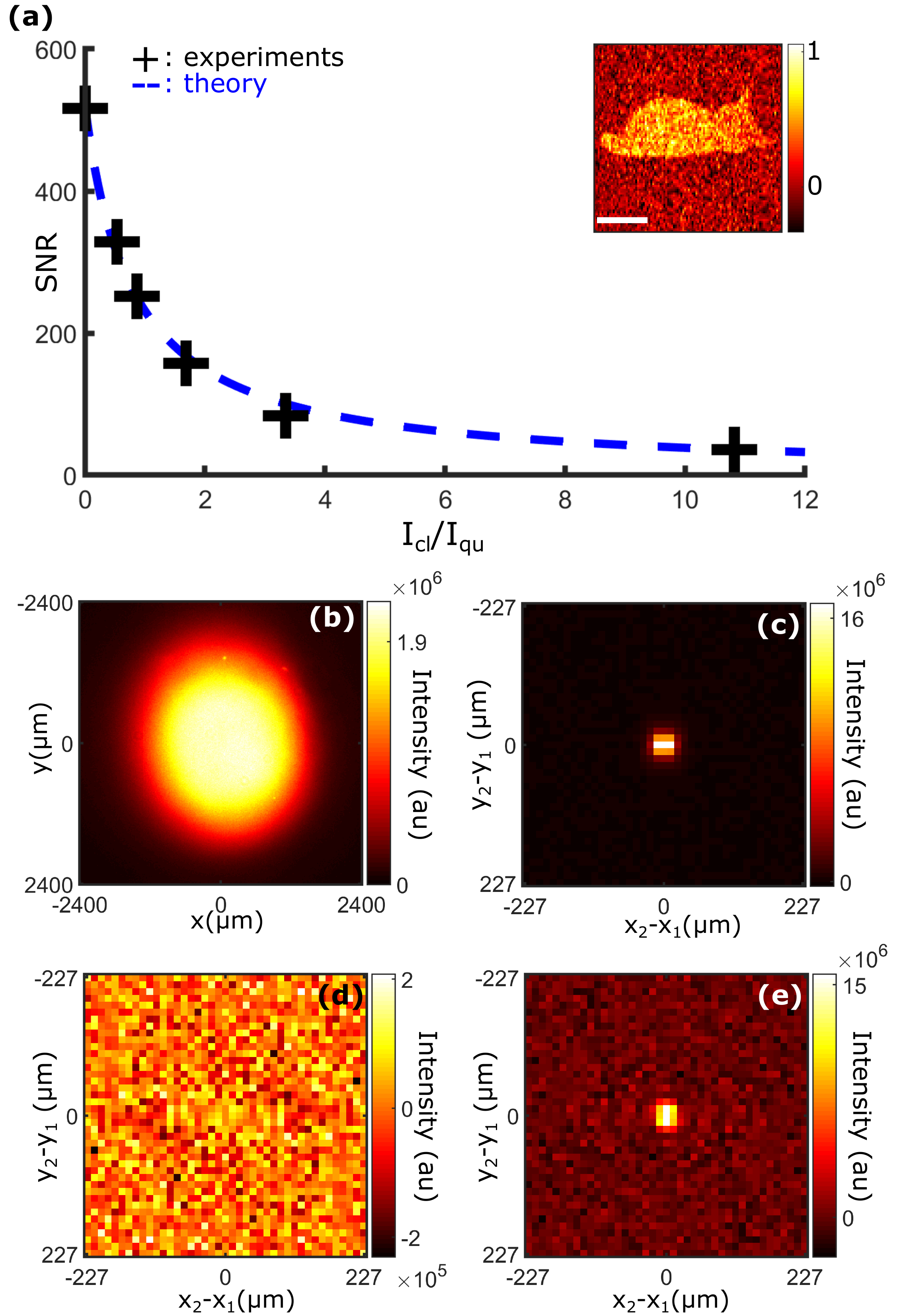}% Here is how to import EPS art
\caption{\label{Figure4} \textbf{Single-to-noise ratio (SNR) in quantum distilled images.} \textbf{(a)} SNRs are represented as function of average intensity ratio between classical and quantum light $ I_{cl} / I_{qu}$ (black crosses) together with a theoretical model (blue dashed line). In this experiment, both sources illuminate homogeneously the camera sensor \textbf{(b)} and SNRs are measured by dividing the peak intensity by the standard deviation of the noise in the minus-coordinate projections of $\Gamma$. \textbf{(b)}, \textbf{(c)} and \textbf{(d)} show minus-coordinate projections acquired for  intensity ratios of respectively $0$, $+ \infty$ and $11$. All experiments are performed by acquiring $N=251600$ images with an exposure time of $\tau = 6$ ms. With these settings, intensity of the quantum source averaged over camera pixels equals $I_{qu} = 939$ grey levels. Inset: normalized quantum image of a `dead cat' reconstructed with an average classical/quantum intensity ratio of $5.5$. White scale bar is $400 \, \mu$m.  }
\end{figure}
These residual single-photon intensities are further investigated by performing a similar experiment using another object $O_3$ (a number `3') that is purposely positioned slightly out of the focal plane of the imaging system. A ground-truth intensity image [Fig.~\ref{Figure3}(a), acquired with the LED turned off] shows the slightly defocused image of $O_3$, well-recognizable by its blurred edges. After turning on the LED, the mixed-intensity image [Fig.~\ref{Figure3}(c)] shows a superposition of the number `3' with a number `6' (object $O_4$). While the number `3' is near-perfectly reconstructed by measuring $\Gamma(\vec{r},\vec{r})$ [Fig.~\ref{Figure3}(d)], we observe again residual intensities in the classical image obtained by image subtraction [Fig.~\ref{Figure3}(e)]. Subtracting this image from the ground-truth of $O_4$ [Fig.~\ref{Figure3}(b), acquired with the photon-pair source turned off] allows us to isolate the residual intensity pattern [Fig.~\ref{Figure3}(f)]. First, we observe that the residual edges of number `3' are thicker than edges of the `dead cat' in Fig.~\ref{Figure2}(c). Indeed, pairs of photons out of the focal plane have a larger correlation width~\cite{reichert_biphoton_2017} and therefore a higher probability that one of them gets blocked by the object. Then, the absence of residual intensity inside the `3' is due to the near-perfect transparency at $810$nm of the printed glass (Thorlabs resolution target). These observations confirm that the residual intensity is not a detection artefact but corresponds to the physical absorption of one photon of a pair when interacting with the object. Since spatial correlations are absent from both single-photon beams and photons emitted by classical light, our intensity-correlation-based approach cannot distinguish between them, preventing us from achieving perfect reconstruction of $O_4$ from the mixed image.\\
While the classical source does not contribute to the intensity-correlation measurement, the presence of uncorrelated photons does reduce the signal-to-noise ratio (SNR) in the measured $\Gamma$. Figure~\ref{Figure4}(a) shows decrease of the SNR with the increase of the average intensities ratio between classical and quantum illumination, $I_{cl} / I_{qu}$, together with its theoretical model (see Methods). In this experiment, the camera is illuminated homogeneously with both quantum and classical light [Fig.~\ref{Figure4}(b)]. SNR values are measured on minus-coordinate projections of $\Gamma$ that represent the probability of detecting two photons from a pair at two pixels separated by a distance $\vec{r_1}-\vec{r_2}$. Fig.~\ref{Figure4} (c) and (e) show minus-coordinate projections of $\Gamma$ acquired respectively at $I_{cl} / I_{qu} =0$ and $I_{cl} / I_{qu} =11$. The central peaks are clear signatures of position correlations between pairs of photons~\cite{moreau_realization_2012,edgar_imaging_2012}. As shown in Fig.~\ref{Figure4}(d), this peak disappears when the camera is illuminated only by classical light i.e.$I_{cl} / I_{qu} = +\infty$. As can be seen, a SNR$>1$ is maintained over a very wide range of classical illumination intensity levels, even when this is 10$\times$ higher than the quantum illumination level, thus indicating that the proposed technique is robust.\\

In conclusion, we have demonstrated the separation of spatial information carried by quantum light (correlated photons) from that carried by classical light (uncorrelated photons) by intensity correlation measurements. For this, we exploited the existence of spatial correlations between pairs of photons generated by SPDC that are absent in classical coherent light. We also showed that the presence of classical light only decreases the quality of reconstructed image but does not change its shape. This novel approach may play an important role for quantum imaging in natural environments, where the object and the camera are contaminated by classical noise or spurious photons.
Moreover, the ability to mix and distinguish information carried by quantum and classical light may have an important impact in quantum communications~\cite{lydersen_hacking_2010}. For example, an image encrypted with correlated photons can be hidden from detectors performing conventional intensity measurements when mixed with a classical image. 
This work paves the way towards the use of mixed light sources composed of both quantum and classical light for improving imaging~\cite{afek_high-noon_2010} and communication technologies~\cite{leonhard_protecting_2018}.

\section*{Methods}

\subsection{Image reconstruction process} The camera is an EMCCD Andor iXon Ultra 897 and was operated at $-60^{\circ}$C, with a horizontal pixel readout rate of $17$Mhz, a vertical pixel shift every $0.3\,\mu$s and a vertical clock amplitude voltage of $+4$V above the factory setting. In each acquisition, $N$ frames are collected with an exposure time $\tau = 6$ ms. No threshold is applied, and all calculations are performed directly using grey values returned by the camera~\cite{defienne_general_2018}. For $\vec{r_2} \neq \vec{r_1}$, $\Gamma(\vec{r_1},\vec{r_2})$ is calculated using the formula:
\begin{equation}
\label{formuleproc}
{\Gamma}(\vec{r_1},\vec{r_2})=\langle I (\vec{r_1}) I(\vec{r_2}) \rangle - \langle I (\vec{r_1}) \rangle \langle I(\vec{r_2}) \rangle
\end{equation}
The first term is the average intensity product:
\begin{equation}
\langle I (\vec{r_1}) I(\vec{r_2}) \rangle = \lim_{N \rightarrow + \infty } \frac{1}{N}\sum_{l=1}^N I_l(\vec{r_1}) I_l(\vec{r_2})
\end{equation}
where $I_l(\vec{r_1})$ [$I_l(\vec{r_2})$] corresponds to the intensity value measured at pixel $\vec{r_1}$ [$\vec{r_2}]$] in the $j^{th}$ frame. Experimentally, this term is estimated by multiplying intensity values in each frame and averaging over a large number of frames (typically $N$ on the order of $10^6-10^7$). Intensity correlations in this term originate from detections of both real coincidence (two photons from the same entangled pair) and accidental coincidence (two photons from different entangled pairs). The second term in equation~\ref{formuleproc} is defined as:
\begin{equation}
\langle I (\vec{r_1}) \rangle \langle I(\vec{r_2}) \rangle = \lim_{N \rightarrow + \infty } \frac{1}{N^2} \sum_{l=1}^{N} \sum_{l'=1}^{N} I_l(\vec{r_1}) I_{l'}(\vec{r_2})
\end{equation}
Experimentally, this term is estimated by multiplying intensity values between successive frames and averaging over a large number of frames:
\begin{equation}
 \langle I (\vec{r_1}) \rangle \langle I(\vec{r_2}) \rangle \approx \frac{1}{N} \sum_{l=1}^{N} I_l(\vec{r_1}) I_{l+1}(\vec{r_2})
\end{equation}
Since there is zero probability for two photons from the same entangled pair to be detected in two different images, intensity correlations in this term originate only from photons from different entangled pairs (accidental coincidence). A subtraction between these two terms (equation~\ref{formuleproc}) leaves only genuine coincidences, which is proportional to the joint probability distribution of photon-pairs. Moreover, the use of intensity products between successive frames, rather than the products of the averaged intensities, allows to reduce artifacts such as spatial distortions in the retrieved $\Gamma$ that are due to fluctuations of the camera amplification gain during the time of an acquisition~\cite{defienne_general_2018}.

Since equation~\ref{formuleproc} is only valid for $\vec{r_2} \neq \vec{r_1}$, diagonal values $\Gamma(\vec{r},\vec{r})$ are approximated to intensity correlation values between neighbouring pixels $\Gamma(\vec{r},\vec{r}) \approx\Gamma(\vec{r},\vec{r} + \delta \vec{r})$, where $\delta \vec{r} = - \delta \, \vec{e_x}$ with  $\delta = 16 \mu m$ and $\vec{e_x}$ is an unit vector. This approximation is justified because the Andor Ultra 897 has a fill factor near to $100 \%$ and the correlation width on the camera is estimated to be $\sigma_r \approx 10 \mu m$~\cite{chan_transverse_2007}.More details about the image reconstruction process are provided in the supplementary document section I and II.

\begin{figure}
\includegraphics[width=0.7 \columnwidth]{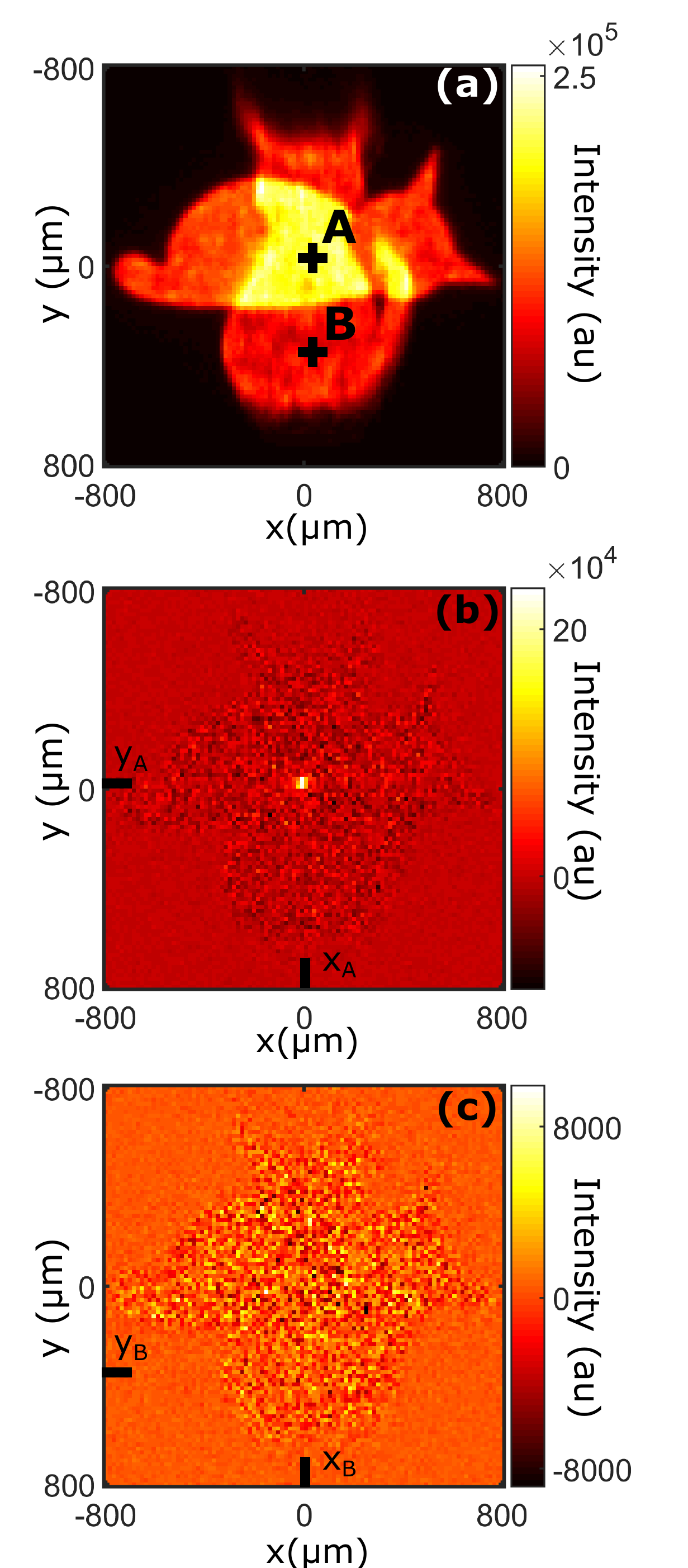}% Here is how to import EPS art
\caption{\label{FigureSM1} \textbf{Conditional projections.} Direct intensity image \textbf{(a)} measured under simulatenous illumination of classical and quantum light. Conditional image $\Gamma(\vec{r}|\vec{A})$ \textbf{(b)} shows an intense peak centered around position $\vec{A}$. Conditional images $\Gamma(\vec{r}|\vec{B})$ \textbf{(c)} is null and flat. }
\end{figure}

A convenient method to visualize $\Gamma$ is to use conditional projections. The conditional projection relative to an arbitrarily chosen position $\vec{A}$, denoted $\Gamma(\vec{r}|\vec{A})$, is an image of intensity correlations between any position $\vec{r}$ and the position $\vec{A}$. For example, two positions $\vec{A}$ and $\vec{B}$ are selected in the direct intensity image in Fig.~\ref{FigureSM1}(a)  and their corresponding conditional projections are shown in Fig.~\ref{FigureSM1}(b) and Fig.~\ref{FigureSM1}(c). $\Gamma(\vec{r}|\vec{A})$ shows an intense peak demonstrating that photon pairs from the SPDC source are transmitted together through the object around position $\vec{A}$. On the contrary, the flat and null pattern of $\Gamma(\vec{r}|\vec{B})$ shows that both photons are absorbed by the object around position $\vec{B}$. \\
\subsection{Signal-to-noise ratio} We define the SNR as the ratio between the central peak intensity and the variance of the noise surrounding it in the minus-coordinates projection of $\Gamma$. This projection is defined as
\begin{equation}
P^{\Gamma}(\vec{r^-})=\int \Gamma(\vec{r},\vec{r}+\vec{r}^-) d \vec{r}
\end{equation}
The SNR formula is derived by adapting the approach described in~\cite{lantz_optimizing_2014}:
\begin{equation}
SNR = \alpha \frac{\sqrt{N} \eta }{2} \left[ 1+\frac{ \sigma_0^2 +   I_{cl}}{\beta ( I_{qu} - \mu_0)} \right]^{-1}
\end{equation}
where $N$ is the number of images acquired, $\eta$ is the quantum efficiency of the camera sensor, and $\mu_0$ and $\sigma_0$ are respectively the camera electronic noise mean value and standard deviation. $\alpha$ and $\beta$ are two parameters that depend respectively on the shape of $\Gamma$ and on the amplification process performed by the camera. $I_{qu}$ and $I_{cl}$ are intensity values of respectively quantum and classical illuminations averaged over all camera pixels, in grey level units (16-bits encoding). In Fig.~\ref{Figure4}, experiments are performed with $N=251600$ and $ I_{qu} = 939$ grey levels (gl). Electrical noise parameters  $\mu_0 = 167 \, gl$ and $\sigma_0 = 32 \, gl $ are estimated independently and $\eta \approx 0.7$ is provided by Andor. Finally, fitting experimental data with the theoretical model (blue dashed curve in Fig.~\ref{Figure4}.a) returns parameters $\alpha= 3.02 \pm 0.22$ and $\beta=0.93 \pm 0.20$ with $R^2 = 0.9955$.\\
\section*{Acknowledgements}
D.F. acknowledges financial support from the UK Engineering and Physical Sciences Research Council (grants EP/M01326X/1 and EP/R030081/1) and from the European Union's Horizon 2020 research and innovation programme under grant agreement No 801060. H.D. aknowledges financial support from the EU Marie-Curie Slowdowska Actions (project 840958).\\
Correspondence and requests for materials should be addressed to hugo.defienne@gmail.com, daniele.faccio@glasgow.ac.uk. HD, MR, JWF, DF conceived and discussed the experiment. HD performed the experiment and analysed the results. HD and MR performed the theoretical analysis. All authors discussed the data and contributed to the manuscript. All authors declare that they have no competing interests.\\

\clearpage

\section*{Supplementary document}

\section{Theory}
\label{theory}
This section provides a brief overview of the theory that underlies our image processing technique. A complete description can be found in~\cite{defienne_general_2018}.\\

We consider the case of a camera illuminated by a source of spatially-entangled photon-pairs, similar to the one shown in Figure 1. Photon-pairs are described by a two-photon wavefunction $\phi(\vec{r_1},\vec{r_2})$, where $\vec{r_1}$ and $\vec{r_2}$ are camera pixels positions. At each camera pixel, photons are converted into intensity values in two steps:
\begin{enumerate}
\item Photons are transformed into photo-electrons by a photo-sensitive screen of quantum efficiency $\eta$
\item Photo-electrons are transformed into intensity values $I_k$ by an amplification register. For $k$ photo-electrons at the input of the register, the camera returns an average grey value that is proportional to $k$: $ I_k = A k + x_0$, where $x_0$ is an electronic noise mean value and $A$ is an amplification parameter.
\end{enumerate}

The camera acquires a set of $N$ images $\{I_l \} _{l \in  [\![ 1,N]\!]}$ using a fixed exposure time. $\langle I(\vec{r}) \rangle$ is defined as the mean intensity value measured at pixel $\vec{r}$:
\begin{equation}
\langle I(\vec{r}) \rangle = \lim_{N \to \infty} \frac{1}{N} \sum_{l=1}^N I_l(\vec{r})
\end{equation}
$\langle I(\vec{r_1}) I(\vec{r_2}) \rangle$ is defined as the mean intensity product value measured between pixels $\vec{r_1}$ and $\vec{r_2}$:
\begin{equation}
\langle I(\vec{r_1}) I(\vec{r_2}) \rangle = \lim_{N \to \infty} \frac{1}{N} \sum_{l=1}^N I_l(\vec{r_1})I_l(\vec{r_2})
\end{equation}

The theoretical analysis is performed under the following assumptions:
\begin{enumerate}[label=\roman*.]
\item Pump laser is operating above threshold to ensure a Poisson distribution of pump photons
\item Pump laser power is low enough to ensure that $>2$ photons generation process in the crystal are negligible
\item Coherence time of photon-pairs is much smaller than the time between two successive images
\item Cross-talk between pixels is negligible
\end{enumerate}

Following the reasoning detailed in the Appendix E of the supplementary document of~\cite{defienne_general_2018}, $\langle I(\vec{r}) \rangle$ and $\langle I(\vec{r_1}) I(\vec{r_2}) \rangle$ are written in function of the camera parameters ($x_0$ and $A$) and the joint probability distribution of photon-pairs $|\phi(\vec{r_1},\vec{r_2})|^2$. On the one hand, $\langle I(\vec{r}) \rangle$ is written as
\begin{equation}
\langle I(\vec{r}) \rangle = x_0 + 2 A \bar{m} \eta P_m(\vec{r})
\end{equation}
where $\bar{m}$ is the mean photon-pair rate and $P_m(\vec{r}) = \int |\phi(\vec{r},\vec{r'})|^2 d \vec{r'}$ is the probability of detecting a photon at pixel $\vec{r}$ (i.e. marginal probability). On the other hand, for $\vec{r_1} \neq \vec{r_2}$, $\langle I(\vec{r_1}) I(\vec{r_2}) \rangle$ is written as
\begin{eqnarray}
\langle I(\vec{r_1}) && I(\vec{r_2}) \rangle = x_0^2 \nonumber \\
 &&+ 2 A x_0 \bar{m} \eta [ P_m(\vec{r_1}) + P_m(\vec{r_2}) ] \nonumber \\
 &&+ 4 A^2 \bar{m}^2 \eta^2  P_m(\vec{r_1}) P_m(\vec{r_2}) \nonumber \\
 &&+ 4 A^2 \bar{m} \eta^2  |\phi(\vec{r_1},\vec{r_2})|^2 
\end{eqnarray}
Finally, the joint probability distribution $|\phi(\vec{r_1},\vec{r_2})|^2$ can be written as
\begin{equation}
\label{equtot2}
|\phi(\vec{r_1},\vec{r_2})|^2 = \frac{\langle I(\vec{r_1}) I(\vec{r_2}) \rangle - \langle I(\vec{r_1}) \rangle \langle I(\vec{r_2}) \rangle}{4 A^2 \bar{m} \eta^2  } 
\end{equation}
While it is commonly thought that photon counting is necessary to compute the joint probability distribution, this result shows that simple operation of a camera without threshold also enables its measurement. 

However, this result is only valid for $\vec{r_1} \neq \vec{r_2}$. As described in Appendix H of the supplementary document of~\cite{defienne_general_2018}, $\langle I(\vec{r})^2 \rangle$ can be written as:
\begin{eqnarray}
\langle I(\vec{r})^2 \rangle &=& 2 A^2 \bar{m} \eta^2 \Gamma(\vec{r},\vec{r}) + 4 A^2 \bar{m} \eta^2 P_m(\vec{r})^2 \nonumber \\
&+& 4 (A^2 + A x_0) \bar{m} \eta P_m(\vec{r}) + \sigma_0 + x_0^2 
\end{eqnarray}
where $\sigma_0$ is the standard deviation of the camera electronic noise. As a result, $ \langle I(\vec{r})^2 \rangle \neq \langle I(\vec{r}) \rangle^2$ and equation (4) is not valid for $\vec{r_1} = \vec{r_2}$. In our experiment, $\Gamma(\vec{r},\vec{r})$ is estimated using the approximation $\Gamma(\vec{r},\vec{r}) = \Gamma(\vec{r},\vec{r} + \delta \vec{r})$, where $\delta \vec{r} = - \delta \, \vec{e_x}$ with  $\delta = 16 \mu m$ (pixel width) and $\vec{e_x}$ is an unit vector.

\section{Measurement of $\Gamma(\vec{r},\vec{r})$}

In our experiment, the camera is an EMCCD Andor Ixon Ultra 897. It was operated at $-60^{\circ}$C, with a horizontal pixel shift readout rate of $17$Mhz, a vertical pixel shift every $0.3\,\mu$s and a vertical clock amplitude voltage of $+4$V above the factory setting. Exposure time is set to $6$ms. All assumptions enumerated in section~\ref{theory} are verified: pump laser operates above threshold with a power of $\sim 50 mW$ [(i) and (ii)], coherent time of the pairs ($\sim 1ps$) is much smaller than the time between two successive frames ($\sim 4 ms$) (iii) and cross-talk between pixels is negligible (iv). In the following, we describe step-by-step the  technique to measure $\Gamma(\vec{r},\vec{r})$:

\begin{enumerate}
\item Acquisition of a set of $N$ images $\{I_l \} _{l \in  [\![ 1,N]\!]}$ at fixed exposure time $\tau=6$ms, with $N$ on the order of $10^{6-7}$. 
\item Estimation of the first term of equation (2) by multiplying pixel values in each image by themselves and averaging over the set:
\begin{equation}
\langle I(\vec{r_1}) I(\vec{r_2})  \rangle \approx \frac{1}{N} \sum_{l=1}^N I_l(\vec{r_1}) I_l(\vec{r_2})  
\end{equation}
in which $\vec{r_1}$ and $\vec{r_2}$ are pixel positions [$\vec{r_1} \neq \vec{r_2}$]. 
\item Estimation of the second term of equation (2) by multiplying pixel values in the $l^{th}$ image by those of the following image $l+1^{th}$ and average over the set:
\begin{equation}
\langle I(\vec{r_1})\rangle \langle I(\vec{r_2})  \rangle \approx \frac{1}{N^2}\sum_{l=1}^{N} I_l(\vec{r_1}) I_{l+1}(\vec{r_2}) \nonumber \\
\end{equation}
By definition, $\langle I(\vec{r_1})\rangle \langle I(\vec{r_2})  \rangle$ equals the limit $N \rightarrow + \infty$ for the following summation:
\begin{eqnarray}
\label{expansion}
&&\frac{1}{N^2} \sum_{l=1}^{N} \sum_{l'=1}^{N} I_l(\vec{r_1}) I_{l'}(\vec{r_2}) = \nonumber \\ 
&& \frac{1}{N^2} \sum_{l=1}^{N} I_l(\vec{r_1}) I_{l}(\vec{r_2}) + \frac{1}{N^2} \sum_{l \neq l'}^{N} I_l(\vec{r_1}) I_{l'}(\vec{r_2})
\end{eqnarray}
 The first term in equation~\ref{expansion} can be written as:
 \begin{equation}
 \frac{1}{N} [\frac{1}{N} \sum_{l=1}^{N} I_l(\vec{r_1}) I_{l}(\vec{r_2})] = o\left(\frac{1}{N}\right)
 \end{equation} 
because by definition of $\langle I(\vec{r_1}) I(\vec{r_2}) \rangle$, the serie $\frac{1}{N} \sum_{l=1}^{N} I_l(\vec{r_1}) I_{l}(\vec{r_2})$. The second term in equation~\ref{expansion} is an estimation of the mean value of intensity product between different frames $\langle I_l(\vec{r_1}) I_{l \neq l'}(\vec{r_2}) \rangle$. Because the probability for two photons of the same pair to be detected in two different frames is null  (coherent time much smaller than camera readout time), intensity values in different frames are independent with each other. In consequence, $\langle I_l(\vec{r_1}) I_{l \neq l'}(\vec{r_2}) \rangle$ can be estimated using only successive frames by calculating the sum $\frac{1}{N^2}\sum_{l=1}^{N} I_l(\vec{r_1}) I_{l+1}(\vec{r_2})$. Experimentally, the use of successive frames to estimate $\langle I(\vec{r_1})\rangle \langle I(\vec{r_2})$ rather than the complete set has the advantage of reducing artifacts as spatial distortions in the measured $\Gamma$, mainly due to fluctuations of the amplification gain of the camera~\cite{defienne_general_2018}.
\\
\item Subtraction between these two terms:
\begin{eqnarray}
&&\Gamma(\vec{r_1},\vec{r_2}) \approx \nonumber \\
&& \frac{1}{N} \sum_{l=1}^N I_l(\vec{r_1}) I_l(\vec{r_2}) -\frac{1}{N^2} \sum_{l \neq l'}^{N} I_l(\vec{r_1}) I_{l'}(\vec{r_2})
\end{eqnarray}
\item As shown in Section I, equation~\ref{equtot2} is only valid for $\vec{r_1} \neq \vec{r_2}$. Estimation of the intensity correlation values $\Gamma(\vec{r},\vec{r})$ from those measured between pixel $\vec{r}=(x,y)$ is then performed using neighbouring pixels $\vec{r'}=(x-\delta,y)$ [$\delta = 16 \mu m =$  pixel size]:
\begin{equation}
\Gamma(\vec{r},\vec{r}) \approx \Gamma((x,y),(x-\delta,y))
\end{equation}
In our experiment, this approximation is valid because the fill factor of the Andor Ixon Ultra is near $100 \%$ and the position correlation width on the camera is estimated from the thickness of the crystal to be $\sigma_r \approx 10 \mu m$~\cite{chan_transverse_2007}.
\end{enumerate} 

\section{Projections of $\Gamma$} 
$\Gamma(\vec{r_1},\vec{r_2}) =\Gamma((x_1,y_1),(x_2,y_2))$ is a 4-dimensional matrix. Its information content can be visualized using two types of projections:
\begin{enumerate}
\item Conditional projection relative to an arbitrarily chosen position $\vec{r'}$, defined as
\begin{equation}
\Gamma(\vec{r} | \vec{r'}) = \frac{\Gamma(\vec{r},\vec{r'})}{\sum_{\vec{r'}} \Gamma(\vec{r} , \vec{r'}) } 
\end{equation}
It represents the probability of detecting a photon from a pair at position ${\vec{r}}$ under the condition that another photon is detected at ${\vec{r'}}$.
\item The minus-coordinate projection, defined as
\begin{equation}
P_-^{\Gamma}({\vec{r}_{-}}) = \sum_{\vec{r}} \Gamma(\vec{r}_{-}+\vec{r},\vec{r})
\end{equation}
It represents the probability for two photons of a pair to be detected in coincidence between pairs of pixels separated by an oriented distance ${\vec{r}_-}$.
\end{enumerate} 

\bibliography{Biblio}

\end{document}